\documentclass[pre]{revtex4}

\usepackage{latexsym}
\usepackage{amsmath}
\usepackage{amssymb}
\usepackage{amsthm}
\usepackage{graphicx}
\usepackage{dcolumn}
\usepackage{bm}

\providecommand{\abs}[1]{\lvert#1\rvert}

\providecommand{\mathsmall}[1]{\mbox{\small{$#1$}}}

\DeclareMathOperator{\erfc}{erfc}

\begin{document}

\title{Quantum Transition State Theory for proton transfer reactions
in enzymes}

\author{Jacques P. Bothma}
\altaffiliation[Current address: ]{Biophysics Graduate Group, University of California, Berkeley, CA 94720-3200 USA}    
\author{Joel B. Gilmore}
\altaffiliation[Current address: ]{ ROAM consulting, 49 Sherwood Road, Toowong Queensland 4066, Australia}             
\author{Ross H. McKenzie}
\email{r.mckenzie@uq.edu.au}
\homepage{condensedconcepts.blogspot.com}

\affiliation{%
The University of Queensland, School of Mathematics and Physics,
Brisbane  4072, Australia}%

\date{\today}

\begin{abstract}
We consider the role of quantum effects in the transfer of hyrogen-like species in enzyme-catalysed reactions. This study is stimulated by claims that the observed magnitude and temperature dependence of 
kinetic isotope effects imply that quantum tunneling below the energy
barrier associated with the transition state significantly
enhances the reaction rate in many enzymes.
We use  a path integral approach which provides a general framework to
understand tunneling in a quantum system which
interacts with an environment  at non-zero temperature.
Here the quantum system is the active site of the enzyme and the environment
is the surrounding protein and water.
Tunneling well below the barrier only occurs for  
temperatures less than a temperature $T_0$ which is determined by
the curvature of potential energy
surface near the top of the barrier. We argue that for most
enzymes this temperature is less than room temperature.
For physically reasonable parameters quantum transition state
theory gives a quantitative description of the temperature dependence
and magnitude of kinetic isotope effects for two classes of
enzymes which have been claimed to exhibit signatures of quantum tunneling.
The only quantum effects are those associated with the transition state,
both reflection at the barrier top and tunneling just below the barrier.
We establish that the friction due to the environment is weak and
only slightly modifies the reaction rate. Furthermore,
at room temperature and for typical energy barriers
environmental degrees of freedom with frequencies much less than
1000 cm$^{-1}$ do not have a significant effect on quantum corrections to the reaction rate.  

\end{abstract}

\maketitle


\section{Introduction}

The possible existence and importance of quantum effects in biomolecular systems is intriguing and controversial.
Whether quantum effects such as
superposition, interference, tunneling,
or entanglement are crucial to the function
of specific biomolecules is receiving 
increasing attention
~\cite{ball,fleming,davies2,helms,Tejero2007,gilmore,reimers,reimers2,toronto,Engel,Allemann}. One might expect most quantum
effects to be destroyed by decoherence\cite{Weiss,slos}
because biomolecules interact strongly with
their ``hot and wet'' environment, i.e.,
they function at room temperature in a highly 
polar solvent, water. Arguably, the most well-established case
 of a quantum effect being crucial for biomolecular function is arguably 
electron tunneling in proteins~\cite{dutton}. Furthermore, it has been argued that by evolution
electron transfer proteins vary and are selected based on tunneling
parameters~\cite{dutton}. The role of tunneling in other biomolecular
systems has also been examined~\cite{devault}.
For example, in myoglobin it has been found that the reaction rate
for binding of carbon monoxide  becomes 
independent of temperature below 80 K,
due to the presence of quantum tunneling~\cite{Alberding1976,Ober1997}.

Over the past two decades the possibility of quantum tunneling of protons in enzymes has attracted considerable attention~\cite{Marcus2006,Marcus2006a,rsoc,RSBKlinman,Allemann}.  The large kinetic isotope
effects and their temperature dependence are
inconsistent with semi-classical        
transition state theory, whereby the chemical reaction
occurs via thermal activation over an energy barrier.
 These discrepancies have been interpreted as evidence
for the presence of tunneling~\cite{Cha1989Science, Grant1989, Bahnson1993, Jonsson1994, Jonsson1996JACS, Nesheim1996, Whittaker1998, Basran1999, Kohen1999Nature, Chowdhury2000, Harris2000, Abad2000, Seymour2002, Angrawal2004,Allemann}.
However, it should be stressed that this evidence
is rather indirect, being based on the values
of fitting parameters for Arrhenius plots for the
temperature dependence of the reaction rate,
where the absolute temperature only varies by
about ten per cent.
In contrast, for chemical reactions involving
much simpler organic molecules, such as
benzoic acid\cite{remmers} or
hydroxymethylene\cite{allen} much more definitive
signatures of proton tunneling have been observed.
These include a temperature independent rate at
low temperatures and tunnel splitting of the ground
state energy~\cite{benderskii,benderskii2}.

Key questions that need to be answered include:

Can some of the experimental results be explained
without invoking tunneling?

To what extent is it necessary to go beyond
semi-classical transition state theory to explain
the observed kinetic isotope effects?

If tunneling does occur, is it important for the function
of the enzyme?

Have enzymes evolved in a manner that enhances
 the contribution of tunneling?

There are currently a wide range of views on the
answers to  these questions.
For example,  a review in Science states that,
``the entire and sole source of the catalytic power
of enzymes is due to the lowering of the 
free energy of activation and any increase in the
generalized transmission co-efficient, as compared to that of the uncatalyzed reaction''~\cite{Garcia-Viloca2004}. 
Villa and Warshel state that,
``the most important contribution to catalysis comes
from the reduction of the activation free energy
by electrostatic effects ... the popular proposal that enzymes
catalyze reactions by special dynamical effects is not supported by a consistent simulation study ...
the interpretation of recent experiments as evidence for
dynamical contributions to catalysis is unjustified.''~\cite{villa}.
In contrast, 
Klinman \textit{et al.} state that, ''Our present findings on hydrogen
transfer under physiological conditions cannot be explained without
invoking both quantum mechanics and enzyme dynamics.''~\cite{Kohen1999Nature}.
Furthermore, Klinman and Kohen proposed that,
``The optimization of enzyme catalysis may entail the evolutionary implementation of chemical strategies that increase the probability of tunneling and thereby accelerate the reaction rate.''\cite{Kohen1998}           
However, Doll, Bender, and Finke\cite{Doll2003,Doll2003a}
synthesized artificial catalysts which performed the same
chemistry in solution (i.e., in the absence of the enzyme)
and exhibited the same kinetic isotopic effects. 
In a paper that focused on simulations 
Schwartz \textit{et al.}~\cite{Caratzoulas2002} express a similar view to Klinman's,
 ``The action of the enzyme
in speeding the chemical reaction, however, is postulated to be
intimately connected to the directed vibrational motion identified
in this paper. Thus, it appears that evolution has designed the
protein matrix of an enzyme not just to hold substrates or
stabilize transition state formation, but rather to channel energy
in a specific chemically relevant direction.''.

Over the past five years several reviews of different
theoretical approaches to this problem have appeared\cite{bert,Marcus2006,Antoniou06,Pu06,Allemann}. 
Most theoretical work makes two particular assumptions which may be debatable, (i) that
the proton transfer process is adiabatic, and (ii) that a single
reaction co-ordinate is adequate. For a detailed discussion of these issues
we refer to a recent review by Marcus.\cite{Marcus2006}
We also note that for non-enzymatic reactions, the first assumption has been brought into question and an alternative  non-adiabatic picture
(analogous to electron transfer) has been proposed.\cite{Peters}
The non-adiabatic proton transfer theory
has been applied to enzymes.\cite{Knapp2002,Meyer05CP,Hatcher2005,Hatcher2007} 
Then the only way for the proton to move from the reactant to product state is via tunneling. The activation energy is then associated with the reorganisation of the environment rather than that of the transition state.
Siebrand and Smedarchina\cite{siebrand} considered
such a approach to explain how some enzymes have a 
large KIE that is weakly temperature dependent.

One approach to examine the role of quantum effects in complex biological molecules is to use quantum mechanical molecular mechanics (QM-MM) simulations. In
this approach atoms which are directly
involved in the reaction are treated quantum mechanically while the rest of the enzyme is treated classically. This approach has been applied to a number of different enzymes\cite{Cui2002,Wang2006JCP,Cui2002JPCB,Alhambra2000,Cristobal2002,Faulder2001,Monard2003,Hammes-Schiffer2004,Tresadern2002,Masgrau2006,Olsson2004}. 
Schwartz and coworkers have as their starting point a Hamiltonian similar
to the one used here~\cite{Caratzoulas2002,Antoniou2002EJBIO,Antoniou06}. 
 They used classical molecular dynamic techniques to simulate specific     reactions and extract the spectral density.

 Using a low-energy effective
Hamiltonian model such as the Caldeira-Leggett Hamiltonian\cite{lego} to capture the essential physics of the relevant process offers a complimentary approach
to QM-MM simulations.
It has the advantage that quantum effects and the role of the environment can
be treated more rigorously, via path integral methods\cite{Weiss,Ankerhold}.
Furthermore, the dependence of behavior on the key physical parameters such
as the curvature of the potential energy surface near the transition state can be elucidated.

In this paper we establish the following points using
a path integral approach.
(i) Tunneling well below the barrier
only occurs for     temperatures less than a temperature
$T_0$ which is largely determined by
the curvature of the top of the barrier. We argue that for most
enzymes this temperature is less than room temperature.
(ii)
For physically reasonable parameters quantum transition state
theory gives a quantitative description of the temperature dependence
and magnitude of kinetic isotope effects for two classes of
enzymes which have been claimed to exhibit signatures of quantum tunneling.
The only quantum effects are those associated with the transition state,
both reflection at the barrier top and tunneling just below the barrier.
(iii) The friction on the proton due to the environment is weak and only
slightly modifies the reaction rate.
(iv)
At room temperature environmental degrees of freedom with frequencies much less than
1000 cm$^{-1}$ do not have a significant effect on quantum corrections to the reaction rate.

\section{Background}\label{Background}

\subsection{Kinetic isotope effects}

The rate coefficient $k_L$, for a chemical reaction
involving a species $L$ at temperature $T$ 
can be written in the Arrhenius form
\begin{equation}\label{Arr}
k_L=A_L \exp{\left(-E_L/(k_B T)\right)},
\end{equation}
where $E_L$ denotes the activation energy for the reaction and $A_L$ is the prefactor. The two quantities $A_L$ and $E_L$ are generally referred to as the Arrhenius parameters. 

The reactions that we will be interested in all involve breaking or forming bonds which contain hydrogen species (protons, deuterium, tritium, hydrogen atoms, and
hydrogen anions). We will only be considering the primary kinetic isotope effects for systems where the hydrogen transfer step is rate limiting.  The kinetic isotope effect (KIE) is generally expressed as the ratio of rate constants $k_{H}/k_{T}$ or $k_{H}/k_{D}$, where the superscripts $P$, $D$ and $T$ denote the reactions in which a proton, deuterium, and tritium are being transferred, respectively.

\subsection{Semi-classical Transition State Theory}

Consider the one-dimensional potential energy shown in Figure 1.
A reaction from $A$ to $C$ proceeds via a transition state at $B$.
The following expression for the rate coefficient is widely used~\cite{ratereview,Eyring1935},
\begin{equation}\label{TST}
k=\kappa\left(\frac{k_B T}{h} \right)\frac{Z^{\ddagger}}{Z_A}\exp{(-E_b/k_B T)} \equiv \kappa \ k_{TST},
\end{equation}
where $Z_A$ is the quantum partition function of the metastable state $A$, and
$Z^{\ddagger}$ is the quantum partition function of the activated complex $B$.
 The parameter $\kappa$ was originally introduced to account for the fact that some trajectories may recross the transition state and return to the reactant state and is also set used to include the effects of quantum tunneling. 
An ad-hoc semi-classical transition state theory the activation energy
is replaced by $E_b - \hbar \omega_0/2$ which corrects for the effect of
the quantum-zero point motion in the reactant well. This leads to kinetic isotope effects because $\omega_0$ depends on the mass of the hydrogenic species
being transferred.

Kim and Kreevoy\cite{kim-kreevoy} gave three criteria that are widely considered to be signatures of quantum tunneling in hydrogen-transfer reactions:
(i) a deuterium kinetic isotope effect $k_H/k_D$
 significantly larger than 6.4 at 20 °C (or 8.9 if secondary isotope effects are included); (ii) an activation energy difference, $E_H-E_D$  larger
than 5.0 kJ/mol; and (iii) a ratio of prefactors, $A_H/A_D$ less than 0.7. 
However, it should be noted that Kim and Kreevoy also stated,
"it appears that completely unambiguous experimental proof that tunneling
occurs at $\sim 300$ K would be impossible to obtain, although tunneling
clearly becomes evident at much lower temperatures."

Following Bell\cite{bellbook}, Kohen {\it et al.}\cite{Kohen2006} state that if tunneling is not significant
prefactor ratios should fall within the range of $0.3 \leq A_H/A_T \leq 1.7$ and $0.5 \leq A_D/A_T \leq 1.4$.         

\begin{figure}[ht]
\centering
\includegraphics[scale=0.6]{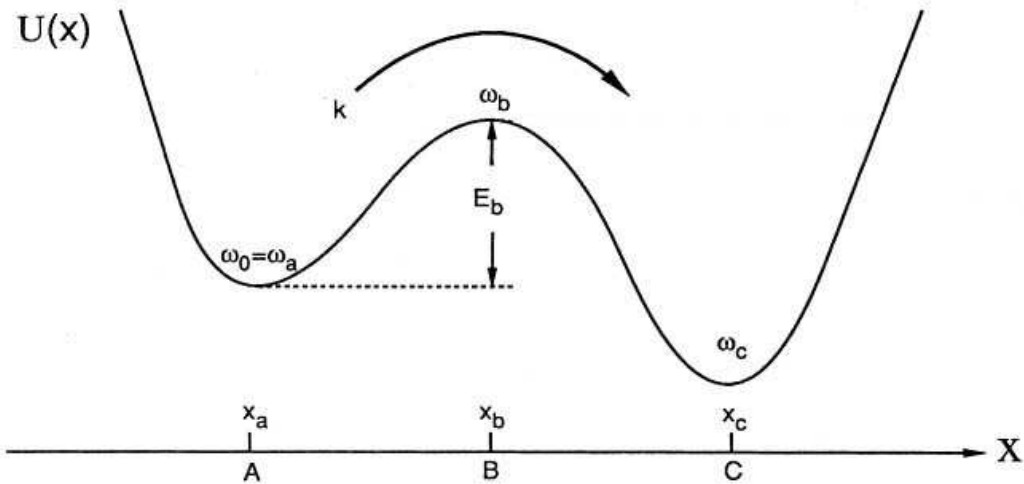}
\caption{Potential energy as a function of the reaction coordinate, $x$, with the metastable reaction state at $A$, the transition state at $B$, and the final product state at $C$. Escape occurs via the forward rate $k$ and $E_b$ is the corresponding activation energy. The angular frequency of oscillations about the reactant state is $\omega_0$, which depends on the curvature of the potential energy surface at the local minimum ($x=x_a$) and the mass of the particle. Similarly the barrier frequency $\omega_b$, depends on the curvature of the potential 
energy at the local maximum ($x=x_b$) and the mass of the particle~\cite{ratereview}. This paper addresses the question as to what extent the reaction $A \to C$ can proceed via quantum tunneling below the barrier for enzyme catalysed hydrogen transfer reactions.}
\label{fig:pot}
\end{figure}

\subsection{Heuristic justification for Quantum Transition State Theory}

Here we reproduce simple arguments described by Weiss\cite{Weiss}. 
A particle in thermodynamic equilibrium in the reactant well A is in a metastable state and so we can think of each quantum state that contributes to the
system partition function $Z$ has an imaginary part, i.e. $\epsilon_n=E_n+i\hbar\Gamma_n$, with $E_n \gg \hbar\Gamma_n$, and
\begin{equation}
\label{partition}
Z = \sum_n \exp(-\epsilon_n/k_BT) \equiv Z_1 +i Z_2 \simeq
 \sum_n \exp(-E_n/k_BT) -      
 i \sum_n \frac{\hbar\Gamma_n}{k_BT}\exp(-E_n/k_BT).       
\end{equation}
The total decay rate out of the reactant well is then
\begin{equation}
\label{decayrate}
k = \frac{1}{Z_1}\sum_n \Gamma_n \exp(-\epsilon_n/k_BT) = \frac{k_B T}{hbar}
\frac{Z_2}{Z_1}.
\end{equation}

If the motion in the reactant well A is described by a single harmonic oscillator in thermal equilibrium, with frequency $\omega_0$, the partition function is,
\begin{equation}\label{TST1}
Z_A =
\frac{1}{\sinh\left(\frac{\hbar \omega_0}{2 k_B T}\right)}.
\end{equation}
If the barrier is an inverted parabola then in this partition
function we can replace $\omega_0$ with $i\omega_b$
to obtain
\begin{equation}\label{TST@}
Z^{\ddagger}
=
\frac{i}{\sin\left(\frac{\hbar \omega_b}{2 k_B T}\right)}
\exp{(-E_b/k_B T)}.
\end{equation}
Assuming no quantum coherence between the bottom and the top
of the barrier the total partition function is then 
$Z=Z_A+i Z^{\ddagger}$.
Substituting this in (\ref{decayrate}) then gives
\begin{equation}\label{qTST}
k=\frac{\omega_b}{4 \pi} 
\frac{\sinh\left(\frac{\hbar \omega_0}{2 k_B T}\right)}
{\sin\left(\frac{\hbar \omega_b}{2 k_B T}\right)}
\exp{(-E_b/k_B T)} 
\end{equation}
An important limitation of this expression is that it
is only well defined for temperatures $T>T_0$ where
\begin{equation}\label{T00}
T_0=\frac{\hbar\omega_b}{2\pi k_B}.          
\end{equation}
The expression (\ref{qTST}) was actually derived by Wigner in 1932\cite{wigner},
using an expression for the energy dependence of the transmission probability
through a parabolic barrier.
This takes into account the fact that in quantum mechanics
a particle with energy $E > E_b$ has less than unit probability
of transmission (i.e. above barrier reflection occurs).
Wigner also assumed  that $T \gg T_0$.     
Bell\cite{bellbook} also derives and discusses this expression.
Note that in the limit that $\hbar \omega_b \ll 2k_B T \ll \hbar \omega_0$, 
(\ref{qTST}) reduces to the semi-classical expression
\begin{equation}\label{scTST}
k=\frac{k_B T}{h} 
\exp{-(E_b-\hbar \omega_0/2)/k_B T)}. 
\end{equation}

In Section \ref{bounce} we use path integral methods
to give a rigorous derivation of Wigner's expression
which will also elucidate its range of validity and the
physical significance of the temperature scale $T_0$.
In Section \ref{Explain} we show that this expression can be used to give a quantitative description of the magnitude and temperature dependence of kinetic isotope effects in several important classes of enzymes.
In Section \ref{divcor} we will see the fact that the rate diverges as $T$ approaches
$T_0$ is an artefact of treating the potential barrier as parabolic.

\subsection{Arrenhius parameters for enzymes
are inconsistent with Semi-classical Transition State Theory}

 Table \ref{tab:KIE} summarises the experimentally determined kinetic parameters for a number of enzymes.           The evidence for tunneling generally comes from examining the prefactor ratios and also the difference in activation energy for different isotopes. Specifically, when these quantities lie outside the
bounds proposed by Bell\cite{bellbook} it is usually claimed that
tunneling occurs. As a reference the activation energy for the different reactions have also been included. The Table also includes some data for other organic reactions  which have kinetic parameters that are inconsistent with semi-classical transition state theory.


\begin{table}[ht]
\caption{Deviations of the Arrhenius
parameters for the kinetic isotopic effects of 
hydrogen transfer reactions in  a range of enzymes from the predictions of semi-classical Transition State Theory. A number of hydrogen transfer reactions involving small organic 
molecules also exhibit parameters that fall outside the semi-classical limits. 
} 
\centering 
\begin{tabular}{|l| c| c| c| c| c|} 
\hline 
\mbox{\hspace{2cm} \textbf{Enzyme, deuterium}} & $k_H/k_D$  &$A_H/A_D$ & $E_D - E_H$  & $E_H$ & Ref. \\ [0.5ex] 
&  (300 K) & & (kJ/mol) &  (kJ/mol)& \\ 
\hline
Semi-classical limits (assuming $\omega_0 \simeq$ 3000 cm$^{-1}$)& $\leq 5$ & 0.5 -1.4 &$\leq$ 3.1&-&~\cite{Kohen2006}\\
\hline
Methylmalonyl-CoA mutase& 35.6 $\pm$ 2.4 & 0.082 $\pm$ 0.028 & 14.3 $\pm$ 0.3 &79 $\pm$ 3 &~\cite{Doll2003}\\
Ethanolamine ammonia lyase & $\sim$30  & 0.038 $\pm$ 2 & 13 $\pm$ 4 & 45 $\pm$ 4  &~\cite{Doll2003a}\\
Soybean lipoxygenase (Wild type)& 81& 18 $\pm$ 5 &3.8 $\pm 0.8$& 8.4 $\pm$ 0.8 &~\cite{Knapp2002}\\
Soybean lipoxygenase, 553V mutant& 82 $\pm$ 6 & 0.3 $\pm$ 0.2 & 10 $\pm$ 2 &10 $\pm$ 2  &~\cite{Meyer2008}\\
Soybean lipoxygenase, 553L mutant& 116$\pm$ 10 & 0.3 $\pm$ 0.4 & 13 $\pm$ 3 & 1.6 $\pm$ 3 &~\cite{Meyer2008}\\
Soybean lipoxygenase, 553A mutant& 93  & 0.12 $\pm$ 0.06 & 16.8 $\pm$ 1.2 & 8.1$\pm$ 0.8 &~\cite{Knapp2002}\\
Soybean lipoxygenase, 553G mutant& 182 $\pm$ 8 & 0.027 $\pm$ 0.034 & 20$\pm$3 & 0.1$\pm$0.1 &~\cite{Meyer2008}\\
Methylamine dehydrogenase  &16.8 &13.3& 0.4 $\pm$ 1.0 & 44.6 $\pm$ 0.5 &~\cite{Basran1999}\\
Methylamine dehydrogenase (Substrate: Ethanolamine) &14.7 &13& 8.4 $\pm$ 1.7 & 43.5 $\pm$ 0.6 &~\cite{Basran2001b}\\
Aromatic amine dehydrogenase (Substrate: Dopamine) &12.9 & 9.4 & 0.7 $\pm$ 0.7 & 50.9 $\pm$ 0.7 &~\cite{Basran2001b}\\
Aromatic amine dehydrogenase (Substrate: Benzylamine) &4.8  & 3.7 & 1.0 $\pm$ 2.3 & 68.1 $\pm$ 1.4 &~\cite{Basran2001b}\\
Aromatic amine dehydrogenase (Substrate: Tryptamine)  & 55 $\pm$ 6 &   & -3.8 $\pm$ 4.6 & 57.3 $\pm$ 3.4 &~\cite{Masgrau2006}\\
Trimethylamine dehydrogenase &4.6 $\pm$ 0.4 & 7.8 $\pm$ 1 & 0.5 $\pm$ 5.2 & 41 & ~\cite{Basran2001}\\
Acyl CoA desaturase &22.9 $\pm$ 2.8 & 19.8 $\pm$ 2.9 & 1.2  & 18.5 &~\cite{Abad2000}\\
Methylamine dehydrogenase  &14.7 & 0.57 & 8.4 $\pm$ 1.1 & 43.5 $\pm$ 0.6 &~\cite{Basran2001b}\\
Peptidylglycine $\alpha$-hydroxylating monooxygenase& 10.4 $\pm$ 0.3 & 5.9 $\pm$ 0.3 & 1.7 $\pm$ 1 & 9  &~\cite{Francisco2002}\\
Sarcosine oxidase&  7.3 & 5 $\pm$ 3 & 0.6 $\pm$ 2.1 & 39.4 $\pm$ 0.9 &~\cite{Harris2000}\\
\textit{E. coli} Dihydrofolate reductase & $4.6 \pm 0.2 $ & $4.0\pm 1.5$ & -0.3 $\pm$ 1 & 12 $\pm$ 1 &~\cite{Sikorski2004}\\
\textit{Thermotoga maritima} Dihydrofolate reductase (25$^{\circ}$C - 65$^{\circ}$C) & 3.3 & $1.56\pm 0.47$ & 2.5 $\pm$ 0.8 & 53.5 $\pm$ 0.4 &~\cite{Maglia2003}\\
\textit{Thermotoga maritima} Dihydrofolate reductase ( $<$ 25$^{\circ}$C) & 5.4 & $0.002 \pm 0.001$ & 19 $\pm$ 5 & 49.9 $\pm$ 1.7 &~\cite{Maglia2003}\\
\hline
\mbox{\hspace{3cm} \textbf{Non-enzyme Reaction}} & &&& & \\
\hline
NpCbl & 35.2 $\pm 1.8$ & 0.14 $\pm$ 0.07 & 12.9 $\pm$ 1.3 & ? &~\cite{Doll2003}\\
AdoCbl & $\sim 29.3$  & 0.16 $\pm$ 0.07 & 12.9 $\pm$ 1.3 & ? &~\cite{Doll2003a}\\
8-MeOAdoCbl & $\sim 21.8 $ & 0.5 $\pm$ 0.4 & 8.8  $\pm$ 2.5 & ? &~\cite{Doll2003a}\\
H$^\bullet$ + c-C$_6$H$_{12}$ $\to$ H$_2$ + c-C$_6$H$^{\bullet}_{11}$ & 9.5 & 0.43 $\pm$ 0.03 & 9.67 $\pm$ 0.25 & - &~\cite{Fujisaki1985}\\
H$^\bullet$ + neo-C$_5$H$_{12}$ $\to$ H$_2$ + neo-C$_5$H$^{\bullet}_{11}$ & 11 & 0.32 $\pm$ 0.04 & 11.00 $\pm$ 0.46 & - &~\cite{Fujisaki1985}\\
H$^\bullet$ + c-C$_6$H$_{12}$ $\to$ H$_2$ + c-C$_6$H$^{\bullet}_{11}$ & 9.5 & 0.43 $\pm$ 0.03 & 9.67 $\pm$ 0.25 & - &~\cite{Fujisaki1985}\\
H$^\bullet$ + n-C$_{10}$H$_{22}$ $\to$ H$_2$ + n-C$_{10}$H$^{\bullet}_{21}$ & 11 & 0.47 $\pm$ 0.03 & 9.41 $\pm$ 0.21 & $\sim$ 30 &~\cite{Fujisaki1983}\\
Proton Transfer in Porphyrin & 11.4 & 0.13& 11.3 & 37.2 &~\cite{Braun1996}\\
Proton Transfer in Porphyrin Anion & 16.5 & 3$\times 10^{-4}$& 25.3 & 17.7 &~\cite{Braun1996}\\
4-nitrophenylnitromethane with tetramethylguanidine & 45 $\pm$ 2 & 0.03 $\pm$ 0.01 & 18 $\pm$ 1 & 17.5
 $\pm$ 0.5 &~\cite{Caldin1973}\\
\hline
\mbox{\hspace{3cm} \textbf{Enzyme, tritium}}  & $k_H/k_T$ & $A_H/A_T$ & $E_T-E_H$  & $E_H$  & Ref.\\
&  (300 K) & & (kJ/mol) &  (kJ/mol)& \\ 
\hline
Semi-classical limits (assuming $\omega_0$ of 3000 cm$^{-1}$)& $\leq 100$ & 0.3-1.7 &$\leq$ 10&-&~\cite{Kohen2006}\\
\hline
Flavoenzyme monoamine oxidase &22 $\pm$ 1 &0.13 $\pm$ 0.03 & 13 & 54 &~\cite{Jonsson1994}\\
\textit{E. coli} Dihydrofolate Reductase & $4.81 \pm 0.06$ & $7.4\pm 0.4$ & -0.4 $\pm$ 1 & 12 $\pm$ 1&~\cite{Sikorski2004}\\
Thymidylate synthase & 7 & 6.8 $\pm$ 2.8  & 0.02 $\pm$ 0.25 & 16 $\pm$ 0.4 & ~\cite{Angrawal2004} \\
Bovine serum amine oxidase & 35& 0.12 $\pm$ 0.04 & 14.2 $\pm$ 0.7& 58 & ~\cite{Grant1989}\\
\hline
\mbox{\hspace{3cm} \textbf{Non-enzyme Reaction}} &  & &  & &  \\ [0.5ex] 
\hline
Porphyrin & 39 & 8$\times 10^{-3}$& 14.3 & 37.2 &~\cite{Braun1996}\\
Porphyrin Anion & 49.6 & 1$\times 10^{-5}$& 37.3 & 17.7 &~\cite{Braun1996}\\
\hline
\end{tabular}
\label{tab:KIE}
\end{table}

\section{Rate Theory}\label{Rate Theory}

\subsection{The Caldeira-Leggett Model Hamiltonian describes
a quantum system interacting with its environment}\label{Model Hamiltonian}

Consider a system which consists of a single particle of mass $M$ described by one degree of freedom and coupled to a large environment which can be represented by a bath of harmonic oscillators. This is equivalent to representing some arbitrary environment in terms of its normal modes. The interaction of the degree of freedom with each of the bath modes is inversely proportional to the volume of the bath. Hence for a macroscopic environment the coupling to each of the individual modes will be weak~\cite{Weiss}.
The  Hamiltonian can be represented as
\begin{equation}
\mathcal{H}=\mathcal{H_S}+\mathcal{H_E}+\mathcal{H_I},
\end{equation}
\noindent
where
\begin{equation}
\mathcal{H_S}=p^2/2M+V(x)
\end{equation}
\noindent
is the Hamiltonian associated with the reaction coordinate $x$. The Hamiltonian of the environment is given by
\begin{equation}
\mathcal{H_E}=\frac{1}{2}\sum_{\alpha=1}^{N}\left(\frac{p^2_{\alpha}}{m_{\alpha}}+m_{\alpha}\omega_{\alpha}^2q_{\alpha}^2\right),
\end{equation}
and describes  $N$ harmonic oscillators where
where $m_{\alpha}$ and $\omega_{\alpha}$ are the mass and frequency of 
the $\alpha$th oscillator. 
 The interaction Hamiltonian is 
\begin{equation}
\mathcal{H_I}=-\sum_{\alpha=1}^{N}F_{\alpha}(x)q_{\alpha}.
\end{equation}
 In our case we will require that the interaction is separable, i.e., $F_{\alpha}(x)=C_{\alpha}F(x)$, and that the dissipation is strictly linear, i.e.,
 $F(x)=x$. This describes state independent dissipation, i.e., the magnitude
of the friction is the same at all points along the reaction coordinate. In that case the complete Hamiltonian is the Caldeira-Leggett Hamiltonian\cite{lego}:
\begin{equation}\label{clasham}
\mathcal{H}=\frac{p^2}{2M}+V(x)+\frac{K_e}{2}x^2 +
\frac{1}{2}\sum_{\alpha=1}^{N}\left[\frac{p_{\alpha}^2}{m_{\alpha}} + m_{\alpha}\omega_{\alpha}^2\left(q_{\alpha}-\frac{C_{\alpha}}{m_\alpha \omega_{\alpha}^2}x\right)^2\right]
\end{equation}
where we have introduced an effective curvature induced by the
environment and defined by
\begin{equation}
K_e    \equiv \sum^N_{\alpha=1} \frac{C_{\alpha}^2}{m_{\alpha} \omega_{\alpha}^2}.
\label{kappaK}
\end{equation}
We also define the effective potential
\begin{equation}\label{veff}
U(x)=V(x)+\frac{K_e}{2}x^2. 
\end{equation}

\subsection{The spectral density describes the frequency-dependent
friction due to the environment}\label{Friction and the spectral density}

The role that the environment plays can be embodied in a single function. It depends on how the coupling strength to each oscillator mode changes with the frequency of the oscillator. This can be   expressed in terms of the memory friction kernel, $\gamma(t)$ defined as\cite{Weiss},
\begin{equation}\label{gamma(t)}
\gamma(t)=\frac{1}{M}\sum_{\alpha=1}^N\frac{C_\alpha^2}{m_\alpha \omega_\alpha^2}\cos(\omega_\alpha t).
\end{equation} 
The Laplace transform of the memory friction kernel, $\hat{\gamma}(z)$, is what determines the effect of the environment on the reaction rate,     
\begin{equation} \label{gam}
\hat{\gamma}(z)=\frac{1}{M} \sum_{\alpha=1}^{N}\frac{C_\alpha^2}{m_\alpha\omega_\alpha^2}\left[\frac{z}{z^2+\omega_\alpha^2}\right].
\end{equation}
The spectral density, $J(\omega)$ is defined as
\begin{equation}
J(\omega) \equiv \frac{\pi}{2}\sum_{\alpha} \frac{C_{\alpha}^2}{m_{\alpha} \omega_{\alpha}} \delta(\omega - \omega_{\alpha}).
\end{equation}
\noindent
and is an alternative means of characterising the coupling to the environment. The spectral density and the Laplace transform of the memory friction kernel 
are related by the identity,\cite{Weiss}
\begin{equation}
\hat{\gamma}(z)=\frac{z}{M} \frac{2}{\pi}\int_{0}^{\infty} d\omega \frac{J(\omega)}{\omega}\frac{1}{\omega^2+z^2}.
\end{equation}
From this we can obtain an upper bound for the friction kernel
\begin{equation}
\label{bound}
\hat{\gamma}(z) \leq \frac{2}{\pi M z} \int_{0}^{\infty} d\omega 
\frac{J(\omega)}{\omega}
=\frac{2 K_e}{\pi M z} 
\end{equation}
where $ K_e$ is the curvature (\ref{kappaK}).
This expression will be used in Section \ref{Estimates} to estimate
the magnitude of the friction.

The simplest kind of dissipation is memoryless friction, $\hat{\gamma}(z)=\gamma$ or $J(\omega)=M\gamma\omega$. However, there is always some microscopic memory which sets the time scale on which inertial effects in the bath are significant. 
The simplest form of damping kernel that captures this 
 is the Drude regularisation,\cite{Weiss} 
\begin{equation}
\begin{split}
\hat{\gamma}(z)&=\frac{\gamma}{1+z/\omega_D} \\ \label{lorfric}
J(\omega)&=\frac{M\gamma\omega}{1+\omega^2/\omega_D^2}.
\end{split}
\end{equation}

We can model a peak in the spectral density
at a frequency $\omega_r$ by
\begin{equation}
{\rm Re} \ \gamma(\omega)= \frac{J(\omega)}{M \omega}
=\frac{\gamma_r (\omega \Gamma)^2}{(\omega^2 -{\omega_r}^2)^2
+ (\omega \Gamma)^2}
\label{gamma-peak}
\end{equation}
which has a value of $\gamma_r$ at the peak which has
a width $\Gamma$.
The corresponding friction kernel is
\begin{equation}
\hat{\gamma}(z)= 
\frac{\gamma_r z \Gamma}{z^2 +{\omega_r}^2 + z \Gamma}.
\label{gamma-peakz}
\end{equation}

\subsection{Classical Kramers Theory defines an effective
barrier frequency}\label{clas kramers}

We first review results for the classical limit of the Hamiltonian.
After averaging over all the environmental variables one finds from Eq. 
\eqref{clasham} the generalised Langevin equation~\cite{Zwanzig1973JStatP}

\begin{equation}
M\ddot{x} +\frac{\partial U}{\partial x} + M \int_0^t\gamma(t-s)\dot{x}(s)ds=\xi(t),
\end{equation}
\noindent
where $\xi(t)$ is the random force the particle experiences and $\gamma(t)$ is the friction kernel that describes the dissipative interaction with the environment. When the total system is prepared initially in thermal equilibrium, the random force $\xi(t)$ becomes a stationary Gaussian noise of vanishing mean, i.e. $\langle \xi(t) \rangle =0$. The classical fluctuation-dissipation theorem gives
\begin{equation}
\langle \xi(t)\xi(0) \rangle =M k_B T  \gamma(t),
\end{equation}
\noindent
where the friction kernel $\gamma(t)$ is defined in Eq. \eqref{gamma(t)}. From this one can then perform a normal mode analysis to evaluate the partition functions entering the transition rate expression~\cite{Weiss}. 
Close to the bottom of the potential well
\begin{equation}
U(x) \approx \mathsmall{\frac{1}{2}}M\omega_0^2(x-x_0)^2, 
\end{equation}
\noindent
and at the barrier top 
\begin{equation}
 U(x) \approx E_b -\mathsmall{\frac{1}{2}} M\omega_b^2 (x-x_b)^2.
\end{equation}
This yields the classical rate which incorporates a dissipative interaction with the environment~\cite{ratereview}
\begin{equation}\label{cmb}
k_{cl} = \frac{\mu}{\omega_b}\frac{\omega_0}{2\pi}\exp(-E_b/k_B T),
\end{equation}
\noindent
and here the effective barrier frequency $\mu$ is the solution of the equation,
\begin{equation}\label{mmuu}
\mu =\sqrt{\frac{\hat{\gamma}^2(\mu)}{4} +  \omega_b^2} -\frac{\hat{\gamma}(\mu)}{2}. 
\end{equation}

In this framework the activation energy has no mass dependence and hence remains unchanged by an isotopic substitution. The only quantities that are altered are $\omega_0$, $\omega_b$ and $\mu$, which all appear in the prefactor. The particle mass term appearing in $\omega_0$ and $\omega_b$ cancel each other out so that the only mass dependence lies in the effective barrier frequency.
 The entire KIE comes from the effective barrier frequency. 
Eq. \eqref{mmuu} then gives the bounds  on the KIE. In the case where one is comparing the rate of a reaction where protium is transferred with a reaction where tritium is transferred this predicts that $1 \leq k_H/k_T \leq 1.7$. Similarly for protium and deuterium $1 \leq k_H/k_D \leq 1.4$. Experimentally the KIE does often depend on the temperature and in the systems of interest often falls outside of the former bounds (See Table \ref{tab:KIE}). The inconsistency between the KIEs predicted by classical Kramers theory and those measured by experiment shows that this classical description does not capture the relevant physics.

\subsection{Path integral representation of Quantum Kramers Theory}\label{Quantum Kramers Theory}

The general problem of quantum tunneling at non-zero temperature  in the
presence of an environment can be treated using
complex-time path integrals~\cite{ratereview,Weiss,Ankerhold}.
Consider the partition function
$Z=\mbox{Tr} \lbrace \exp(-\mathcal{H}/k_B T) \rbrace$,
where $\mathcal{H}$ denotes the full Hamiltonian operator corresponding to the system plus environment. This quantity can be expressed in the form of a functional path integral over the tunneling coordinate $x(\tau)$~\cite{ratereview},
\begin{equation}\label{fint}
Z=\int \mathcal{D}x(\tau)\exp\lbrace -S_E [ x(\tau)]/\hbar \rbrace,
\end{equation}
where $\tau = i\mbox{t}$ is a real variable. This integral sums
over all paths $x(\tau)$ that satisfy the periodic boundary condition
\begin{equation}
\label{periodic}
x(\tau=-\theta/2)=x(\tau=\theta/2).
\end{equation}
 with period $\theta$ determined by the temperature,
\begin{equation}
\label{theta}
\theta =\hbar/(k_B T).
\end{equation}
  After integrating over the bath modes the effective Euclidean action takes the form,
\begin{equation}\label{action}
S_E[x]  =\int^{\theta /2}_{-\theta /2} d\tau\lbrace \frac{M}{2} \dot{x}^2(\tau) +U[x(\tau)]\rbrace\\
 + \frac{1}{2} \int^{\theta /2}_{-\theta /2} d\tau \int^{\theta /2}_{-\theta /2} d\tau' \zeta(\tau -\tau')x(\tau)x(\tau').
\end{equation}
The influence kernel $\zeta(\tau)$ is periodic in imaginary time with period $\theta$. It is
related to 
$\hat{\gamma}(z)$,   the Laplace transform of the memory friction (see Eq. \eqref{gam}) and
can be represented in terms of a Fourier series as~\cite{Grabert87PRB}
\begin{equation}\label{ker}
\zeta(\tau) =\frac{M}{\theta} \sum_{n=-\infty}^{\infty} \abs{\nu_n} \hat{\gamma}(\abs{\nu_n})\exp(i\nu_n \tau)
\end{equation}
satisfying
\begin{equation}\label{norm}
\int^{\theta /2}_{-\theta /2} \zeta(\tau) d\tau=0,
\end{equation}
and where  $\nu_n$ are the Matsubara frequencies for bosons,
\begin{equation}\label{norm2}
\nu_n=n 2\pi k_B T/\hbar.
\end{equation}

For a metastable potential the partition function has an imaginary part which
can be related to the escape rate from the potential\cite{Weiss}.
  The dominant contributions to the partition function, and indirectly the rate expression, come from the vicinity of paths in which the action 
\eqref{action} is stationary. These paths, $x_e(\tau)$, satisfy
the equation of motion
\begin{equation}\label{EL}
M\ddot{x_e}(\tau)-\frac{\partial U[x_e(\tau)]}{\partial x_e(\tau)}-\int_{-\theta/2}^{\theta/2}d\tau'\zeta(\tau-\tau')x_e(\tau') =0,
\end{equation}
and the periodic boundary condition
(\ref{periodic}).
In the absence of dissipation, i.e. $\zeta(\tau)=0$, the evolution of $x_e(\tau)$ in imaginary time corresponds to real-time motion in the metastable inverted potential $-U(x)$.\cite{miller75,MillerACP} 

Because of Eq. \eqref{norm}, Eq. \eqref{EL} has two trivial but
physically important solutions. The first where the particle remains at the bottom of the reactant well ($x_e(\tau)=x_a$) and the other where it sits on top of the barrier ($x_e(\tau)=x_b$).
The latter corresponds to thermal activation over the barrier top.

\subsection{The bounce solution describes quantum tunneling which
only occurs below  a temperature, $T_0$}\label{bounce}

A non-trivial periodic solution 
to Eqn. (\ref{EL}) (which has been dubbed the bounce or instanton 
solution) describes quantum tunneling below the barrier.\cite{miller75,Coleman} 
{\it This solution only exists below a temperature} $T_0$~\cite{ratereview}.
In the absence of dissipation, an analytic solution has been found
for an inverted parabola, an Eckart potential,\cite{MillerACP} and a cubic potential.\cite{Freidkin}
In the presence of Ohmic dissipation, an analytic solution
for a cubic potential has been found for specific values of the dissipation.\cite{Riseborough}

 For temperatures $T>T_0$ the period of the $\theta$-periodic orbit is not of sufficient length to admit an oscillation of the particle in the classically forbidden regime.
For the case of zero temperature the bounce
solution gives a rate which is directly proportional
to the tunneling probability calculated from the WKB approximation,
\begin{equation}
T(E) = \exp \left( -2S(E)/\hbar \right)
\end{equation}
where $S(E) $  is the value of classical action along
the imaginary time trajectory,
\begin{equation}
S(E) = \sqrt{2 M} \int_{x_{1}}^{x_{2}} dx
\left[ U(x)  - E \right]^{1/2}, 
\end{equation}
where $x_1$ and $x_2$ are the classical turning points for energy $E$. 

For temperatures $T>T_0$ the bounce solution does not exist
and the only contribution to the
path integral comes from the constant solution ($x_e(\tau)=x_b$) where the particle sits at the barrier top. 
 Fig. 40 in Ref. \onlinecite{ratereview} depicts the different kinetic regimes which occur for different temperatures.


We now focus on the case of  a parabolic barrier.
In the absence of any dissipation  ($\gamma=0$), 
the crossover temperature has the value given by (\ref{T00}). 
In the presence of dissipation,
the crossover temperature $T_0$ is given by   
\begin{equation}\label{T_0}
T_0 \equiv \frac{\hbar\mu}{2\pi k_B} = 0.23 \ {\rm  K}\frac{\mu}{{\rm cm}^{-1}},
\end{equation}
where $\mu$ is the effective barrier frequency defined in Eq. \eqref{mmuu}.
This means that tunneling can only occur at room temperature
if $\mu > 1300$ cm$^{-1}$.
In an appendix it is shown how for a Lorentzian spectral
density $\mu$ is reduced by friction.

\section{Quantum Correction Factor: $T >   T_0$}
\label{Quantum Correction Factor: TggT0}

In the high-temperature regime for a parabolic barrier
 one can obtain an       analytic expression for the rate constant~\cite{Wolynes1981PRL,Hanggi1987ZPhysB},
\begin{equation}\label{9.32}
k(T)=k_{cl} c_{qm} \equiv \biggl[\frac{\mu}{\omega_b}\biggl(\frac{\omega_0}{2\pi}\biggr)\exp(-E_b/k_BT)\biggr] 
\left\lbrace\hspace{1.5mm}\prod_{n=1}^{\infty}\frac{\omega_0^2+n^2\nu^2+n\nu\hat{\gamma}(n\nu)}{-\omega_b^2+n^2\nu^2+n\nu\hat{\gamma}(n\nu)}\right\rbrace
\end{equation}
The first term in the square brackets denotes the classical Kramers rate for memory friction (Eq. \ref{cmb}). Here $\nu$ is the smallest
Matsubara frequency,
\begin{equation}\label{norm3}
\nu= 2\pi k_B T/\hbar 
\end{equation}
and 
we must have $\nu > \mu$ where $\mu$ is the effective barrier frequency
given by (\ref{mmuu}).
We note that (\ref{9.32}) is proportional to $Z^\ddagger/Z_A$ where the corresponding partition functions are for damped harmonic oscillators.\cite{Weiss}

The quantum correction to the rate expression, $c_{qm}$, is encapsulated by the term inside the  curly brackets. For $T \gg T_0$ this correction factor approaches unity. Moreover, it always exceeds unity which implies that quantum effects always \textit{enhance} the classical rate. 

\subsection{The weak friction limit reduces to Wigner's expression}\label{FL}

This limit has to be treated with some care as when the friction is exceptionally weak thermal equilibrium no longer prevails in the reactant well. It has been shown that so long as the following condition is satisfied the assumption can be made that the reactant system is always in thermal equilibrium \cite{ratereview},
\begin{equation}
\frac{\hat{\gamma}(\mu)}{\omega_b} >\frac{k_B T}{E_b}.
\end{equation}
However, we note that for some of the enzymes shown in  Table I
the activation energy is sufficiently small this assumption may not
be justified.

In the limit where $\gamma \to 0$ Eq. \eqref{9.32} can be simplified such that the correction factor can be written as
\begin{equation}
\label{ccqqmm}
c_{qm}=\frac{\omega_b}{\omega_0} \frac{\sinh\left(\frac{\hbar \omega_0}{2 k_B T}\right)}{\sin\left(\frac{\hbar \omega_b}{2 k_B T}\right)}.
\end{equation}
This expression will be a reasonable approximation provided
that for all $n=1,2,..$
\begin{equation}\label{weak}
\frac{\hat{\gamma}(n \nu)}{n \nu} \ll 1.                
\end{equation}
Since $\nu \sim 1300$ cm$^{-1}$ at room temperature this means that
any friction associated with environmental
modes of much lower frequency may have little effect
on the quantum correction factor.

\begin{center}
\begin{figure}[ht]
	\centering
		\includegraphics[scale = 0.9]{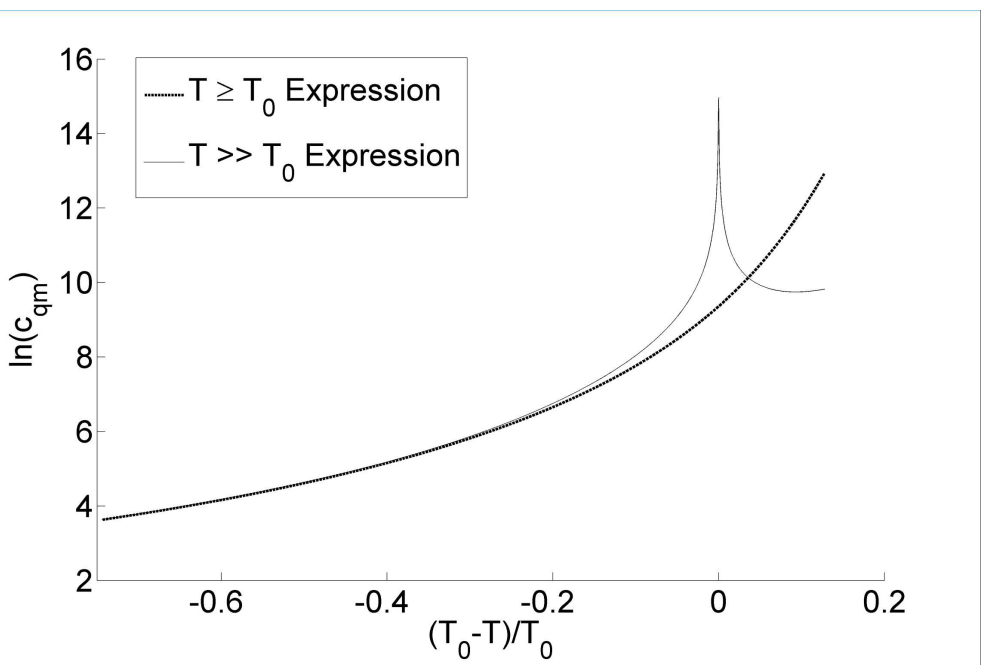}
	\caption{Comparison of the approximate quantum correction factor
with the exact result. The approximate expression does not take into account
the non-parabolic character of the barrier top.
For temperatures above about $1.1T_0$ the two expressions (\ref{ccqqmm}) and
(\ref{qMT}) are in good agreement. However as $T_0$ is approached the high temperature expression diverges.  The parameter values used in generating these plots are: $\omega_0$ = 3000 cm$^{-1}$, $\omega_b$ = 1000 cm$^{-1}$, $ \kappa(T_0)$ = 10. This shows that we are justified in using our expression (\ref{QTSTFit})
 to describe the kinetic isotopic effect
  provided that the crossover temperature $T_0$ is less than about 250 K.
}
	\label{fig:comppp}
\end{figure}
\end{center}

Note that the expression (\ref{ccqqmm}) diverges as $T \to T_0^+$.
This divergence turns out to be
an artefact from treating the potential near the transition state as a perfect inverted parabola. Below we will show how a more a rigorous
treatment shows  that in this temperature regime the correction factor is always finite. Furthermore, for realistic parameters (\ref{9.32}) 
is a good approximation down to $1.1T_0$.
(See Fig. \ref{fig:comppp}).




\subsection{Apparent Arrhenius parameters in the weak friction limit} \label{APP}

 Most kinetic experiments are performed over a narrow temperature range.
 The temperature dependence appears to be activated (i.e, a plot of $\ln(k_L/k_T)$ vs. $1/T$ is linear over the measured temperature range) and so it is natural to determine the Arrhenius parameters.
If the full QTST rate expression is expanded about room temperature one can obtain an expression for the KIE that has a simple activated temperature dependence.  The subscripts $L$ and $L^{\dagger}$ denote possible combinations of the three different isotopes of hydrogen.  This is typically what is done in experiments where the heavier isotopes are used as a reference. Combining the results from Eq. \eqref{9.32}  and Eq. \eqref{ccqqmm} gives an expression
for the rate constant.
The KIE is given by 
\begin{equation}
\frac{k_L}{k_{L^{\dagger}}}=\sqrt{\frac{m_{L^{\dagger}}}{m_L}}
\frac{\sinh\left(\frac{\hbar \omega_0}{\sqrt{4m_{L}} k_B T}\right)}{\sinh\left(\frac{\hbar \omega_0}{\sqrt{4m_{L^{\dagger}}} k_B T}\right)}
\frac{\sin\left(\frac{\hbar \omega_b}{\sqrt{4m_{L^{\dagger}}} k_B T}\right)}
{\sin\left(\frac{\hbar \omega_b}{\sqrt{4m_L}) k_B T}\right)}
	\label{eq:KIE}
\end{equation}
\noindent
where $m_L$ is the unitless mass number of the $L$ isotope and $\omega_0$ and $\omega_b$ are the ground state oscillation frequency and barrier frequency, respectively for hydrogen.  For a typical C-H 
stretch frequency, $\hbar\omega_0 \gg 2k_B T$, at room temperature so the hyperbolic sine terms can be approximated as exponential functions. Over the biologically relevant temperature range $T$ only varies by less than 10\% and so one can expand the other temperature dependent parts of the expression up to linear terms in $1/T$. This gives an expression for the KIE which has a simple activated temperature dependence with the following apparent Arrhenius parameters.

\begin{equation}\label{SCUL}
\begin{split}
\frac{A_L}{A_{L^{\dagger}}} &= \sqrt{\frac{m_{L^{\dagger}}}{m_L}}\frac{\sin(\beta_R\hbar\omega_b/\sqrt{4m_{L^{\dagger}}})}{\sin(\beta_R\hbar\omega_b/\sqrt{4m_L})}\\
\times & \exp\left[-\frac{\beta_R\hbar\omega_b}{2}\left(\frac{\cot(\beta_R\hbar\omega_b/\sqrt{4m_{L^{\dagger}}})}{\sqrt{m_{L^{\dagger}}}}-\frac{\cot(\beta_R\hbar\omega_b/\sqrt{4m_L})}{\sqrt{m_L}}\right)\right].\\
E_{L^{\dagger}}-E_L&=\frac{\hbar\omega_0}{2}\left(\frac{1}{\sqrt{m_L}}-\frac{1}{\sqrt{m_{L^{\dagger}}}}\right)\\
+ & \frac{\hbar\omega_b}{2}\left(\frac{\cot(\beta_R\hbar\omega_b/\sqrt{4m_{L^{\dagger}}})}{\sqrt{m_{L^{\dagger}}}}-\frac{\cot(\beta_R\hbar\omega_b/\sqrt{4m_L})}{\sqrt{m_L}}\right),
\end{split}
\end{equation}
where $T_R = 1/k_B\beta_R$ is the temperature around which the expansion is performed. From these expressions it is also possible to place bounds on the apparent Arrhenius Parameters. When $T_R \geq T_0$ and $m_L < m_{L^{\dagger}}$ it can be shown that the Arrhenius parameters are monotonic functions of $\beta_R\hbar\omega_b$. 

On their own the expressions in Eq. \eqref{SCUL} may not seem to shed much light. However, evaluating these expressions for typical parameter values  shows that typical values for $\omega_0$ and $\omega_b$ give kinetic parameter values that are inconsistent with the predictions of standard semi-classical rate theory. Moreover, Table \ref{tab:apar} shows that the parameter trends are consistent with experimentally determined Arrhenius parameters for a number of systems. In a number of systems where tunneling has been invoked the difference in apparent activation for the different isotopes is 
greater than would be predicted by semi-classical theories. The effective activation energies derived from QTST are quantitatively similar to a number of the experimental values and significantly exceed the semi-classical values.
The prefactor values that have been observed experimentally are smaller than would be expected from semi-classical theories.

\begin{table}[ht]

\caption{The effective Arrhenius parameters calculated for the expanded QTST expression are compared with experimentally determined values. Both the calculated and experimentally determined parameters are inconsistent with a standard semi-classical analysis. This shows that there are a subset of enzymatic systems where the anomalous kinetics can be explained by Quantum Transition State Theory.}
\centering 
\begin{tabular}{|l| c| c| c| c|} 
\hline  
\mbox{\hspace{0cm} \textbf{Parameter Values/Experimental System}} & $E_T-E_H$ (kJ/mol) &  $A_H/A_T$ & $E_T-E_D$ (kJ/mol) &$A_D/A_T$  \\ 
\hline
Semiclassical limits  & $\leq 10.1$ & 0.3-1.7 & $\leq 3.1$ & 0.5-1.4 \\
$ \omega_0 =3000 $ cm$^{-1}$, $ \omega_b =1000 $ cm$^{-1}$, $ T_R = 288 $K & 16& 0.08& 3.6& 0.7 \\
Bovine Serum Amine Oxidase \cite{Grant1989}& 14.2 $\pm$ 0.7& 0.12 $\pm$ 0.04 & 4.51 $\pm$ 0.48& 0.51 $\pm$ 0.1 \\
Flavoenzyme Monoamine Oxidase B \cite{Jonsson1994}& 13 & 0.13 & 4.1 & 0.51  \\
\hline
\end{tabular}
\label{tab:apar}
\end{table}

\section{Non-parabolicity of the barrier only matters at
temperatures close to $T_0$}
\label{divcor}

In the regime where $T \sim T_0$ the action associated with the bounce solution and the trivial solution at the top of
the barrier become comparable (See Section \ref{bounce}).  
In the limit where $T \to T_0$ the expression for the quantum correction given in Eq.\eqref{9.32} diverges.
 This divergence is a consequence of the fact that the saddle point approximation of the imaginary time functional integral employed in evaluating the expressions presented so far breaks down in the vicinity of $T_0$. 
In order to obtain an expression which is valid around the crossover temperature the analysis needs to be extended to include
the effects of a non-parabolic barrier\cite{Grabert87PRB}.
 The potential is expanded around the barrier top to give
\begin{equation}
V(x)=E_b-\frac{1}{2}M\omega_b^2(x-x_b)^2+\sum_{k=3}^\infty \frac{M c_k (x-x_b)^k}{k}.
\end{equation}

In the weak friction limit (see Section \ref{FL})  the rate expression can be simplified to give a quantum correction factor which takes the form  
\begin{equation}\label{qMT}
\begin{split}
c_{qMT}&=\frac{\omega_b}{\omega_0}\frac{\sinh(\hbar\omega_0/2k_B T)}{\sin(\hbar\omega_b/2k_B T)}\\
   \hspace{3mm} &  \times\sqrt{\pi}(-\epsilon)(1-\epsilon/2)\kappa\erfc(-\epsilon(1-\epsilon/2)\kappa)
\exp(\epsilon^2(1-\epsilon/2)^2\kappa^2).
\end{split}
\end{equation}
where
\begin{equation}
\epsilon \equiv (T_0-T)/T_0. 
\end{equation}
 This extension  is only required when the temperature is close to the crossover temperature. The extent of this crossover region depends on both the scaled temperature $\epsilon$, and the parameter $\kappa$,  defined as
\begin{equation}
\kappa = \omega_b^2 \sqrt{\frac{8M}{B k_B T_0}}
\end{equation}
where
\begin{equation}
\label{exactcq}
B= 4c_3^2/3\omega_b^2 + 3c_4.
\end{equation}
If the barrier is high and the potential reasonably smooth then $M \omega_b^2/B$ is an energy which is of the order of the barrier height so that $\kappa$ is of the order of $\sqrt{(E_b/\hbar\omega_b)}\gg 1$\cite{Grabert87PRB}.
However, caution is in order because
for $E_b \sim$ 50 kJ/mol $\sim  4000$  cm$^{-1}$ this condition is only
weakly satisfied.
Also, quantum chemistry calculations discussed in 
Section \ref{omegabest} show highly non-parabolic barriers.

\section{Estimates of model parameter values for enzymes}\label{Estimates}

\subsection{Frequency of Oscillations in the Reactant Well ($\omega_0$)}

The parameter $\omega_0$ is defined as $\sqrt{U''(x_0)/M}$, where $M$ is the effective mass of the hydrogen isotope being transferred and $U''(x_0)$ is the curvature of the potential at $x=x_0$. Using IR spectroscopy the oscillation frequencies of the different chemical bond stretches can be measured experimentally~\cite{Settle1997}. Results from molecular dynamics simulation show that the value of $\omega_0$ is similar to the value obtained from IR spectroscopy. In~\cite{Alhambra2000} it was established that the oscillation frequency of the C-H bond that participates in the reaction, $\omega_0$ is 2900 cm$^{-1}$. A typical IR spectrum of an organic molecule would contain peaks corresponding to C-H stretch frequencies in the range  2700-3300 cm$^{-1}$~\cite{Settle1997}.

The value of $\omega_0$ has the potential to be influenced by interactions with atoms in the active site. This interaction has been known to decrease the stiffness of the bond~\cite{Kotting2004}. As a result it is possible that in some of the systems we are studying the binding of the substrate could bring about a change in the ground state oscillation frequency, but based on what happens with 
hydrogen bonds it is unlikely this could ever be a reduction larger than a factor of two. 

\subsection{Barrier Frequency ($\omega_b$)}
\label{omegabest}

The barrier frequency, $\omega_b$, strongly influences to what extent quantum effects affect the kinetics of the hydrogen transfer process. It depends on both the curvature of the potential at the barrier and also on the mass of the hydrogen species being transferred. It is a parameter which 
would be difficult to obtain experimentally. Generally one needs to resort to quantum chemistry calculations to obtain an estimate of $\omega_b$. For our purposes $\omega_b$ is a crucial parameter. Firstly, it sets the upper limit of the crossover temperature, $T_0$, around which the kinetics becomes more classical in nature. Secondly, in the intermediate and high temperature regime it strongly influences exactly how much the rate gets modified due to quantum corrections. Table \ref{tab:omegab}  shows the values of the barrier frequency that have been calculated using quantum chemistry for a number of different enzymes.

\begin{table}[ht]
\caption{Barrier frequency values
 for several enzymes estimated from quantum chemistry calculations of potential energy surfaces for the hydrogen transfer reaction. The maximum crossover temperature $T_0$ is related to $\omega_b$ by Eqn. (\ref{T00}).
Friction will tend to reduce these values. The corresponding temperatures for deuterium and tritium transfer will be about 30 and 50 per cent lower, respectively. SCC-DFTB denotes Self-Consistent Charge-Density Functional Tight Binding.
AM1 and PM3 are semi-empirical methods, and SRP denotes a Specific Reaction Parametrisation, against a DFT calculation using the B3LYP functional and
the 6-31G* basis set.
}
\centering
\begin{tabular}{|c|c|c|c|c|c}
\hline
Enzyme & Level of theory 
 & {$\omega_b$(cm$^{-1}$)} & Max $T_0$ (K) &Ref.\\
 \hline
Triosephosphate Isomerase (TIM) model (gas phase)& AM1-SRP   & 1365 &  315  &~\cite{Cui2002}\\
TIM model (in water)& AM1-SRP   & 591  &  140  &~\cite{Cui2002}\\
TIM model (in protein model)& AM1-SRP   & 798  &  190  &~\cite{Cui2002}\\
\hline
Liver Alcohol Dehydrogenase(LADH)& SCC-DFTB   & 783 & 180 &~\cite{Cui2002JPCB}\\
LADH &  AM1 & 1046 & 240 &~\cite{Alhambra2000}\\
LADH & AM1 &  1229 & 240 &~\cite{Tresadern2002}\\
\hline
Monoamine Oxidase  & B3LYP/6-31G*  & 1054  & 240 &~\cite{MAOQ}\\
Monoamine Oxidase  & PM3 &   1782 & 410 &~\cite{MAOQ}\\
\hline
Methylamine dehydrogenase (MADH)& PM3 &   2000  & 460 &~\cite{Faulder2001}\\
MADH& AM1-SRP   & 2218  & 510 &~\cite{Tresadern2002}\\
MADH& PM3       & 2000  & 460 &~\cite{Nunez2006}\\
\hline
Soybean Lipoxygenase & PM3/d  & 2913  & 670 &~\cite{Tresadern2002}\\
\hline
Aromatic amine dehydrogenase (tryptamine) & PM3-SRP &  2057  & 450 &~\cite{Masgrau2007}\\
\hline
\end{tabular}
\label{tab:omegab}
\end{table}

From these calculated values of $\omega_b$ it is clear that in a number of the cases considered the barrier frequency is of the order of 1000 cm$^{-1}$, which corresponds to a maximum crossover temperature of around -40 $^{\circ}$C.  Of the enzymes that appear in the table both Methylamine dehydrogenase and Soybean Lipoxygenase have a large KIE, associated with a prefactor ratio which is much larger than one (See Table \ref{tab:KIE}).

 It needs to be kept in mind that the values for the the barrier curvature need to be taken with some caution. Most of the values listed in Table \ref{tab:omegab} were determined using hybrid QM/MM calculations. These techniques encounters some methodological problems which have the potential to strongly influence the calculated barrier frequencies~\cite{Titmuss2000}.

A recent study\cite{Dybala-Defratyka2007} performed a QM-MM study of the 
hydrogen transfer reaction in methylmalonlyl-CoA mutase, treating 44 atoms
near the active site at the AM1 level. They emphasize the role of
 corner cutting but do not give a value for the curvature of the barrier.
The "representative tunneling energy" (the energy at which the
product of the transmission coefficient and the Boltzmann factor is
a maximum) is about 1400 cm$^{-1}$ below the
top of the barrier.

Obtaining a reliable value for the barrier curvature from
computational chemistry represents a major challenge.
We note how the Table shows that 
the values obtained depend on the level of theory.
Other factors to consider are the role of anharmonicity,
dependence of the results on the active site geometry,
and the fact that in a dynamic environment (protein plus water)
there are actually many reaction paths.

\subsection{Frequency-dependent Friction due to the Solvent-Protein Environment}

The analysis in the previous Sections shows that the magnitude and
frequency dependence of friction $\gamma(\omega)$,
and its Laplace transform, the memory friction
kernel $\hat{\gamma}(z)$, can lead to not just
quantitative but also qualitative
differences in the reaction rate.
For example, if  $\hat{\gamma}(\omega_b) > \omega_b$, then 
 the friction can
substantially reduce the temperature $T_0$ (see Figure \ref{T0fri})
below which
the ``bounce'' solution exists, i.e., strong enough friction
can prevent tunneling from the bottom of the well occuring.
The friction also causes vibrational energy relaxation
and dephasing of the vibrations in the reactant well.\cite{May}
The former is proportional to 
${\rm Re} \ \gamma(\omega_0).$
Hence, measurements of the relaxation and dephasing
rates provides a means to determine
the magnitude of the friction.\cite{kuhn}

There are believed to be several main sources of friction
associated with bond deformation and breaking in biomolecules.
The first source is the interaction of the dipole moment associated
with displacement of the proton with the fluctuating electric
field associated with fluctuations in its environment,\cite{ratner}
the surrounding protein and solvent.
Recent ultrafast infrared spectroscopy experiments
of the OH stretch of HOD in liquid D$_2$O have
shown that in bulk water this
fluctuating electric field is the dominant source of vibrational
dephasing.\cite{geissler,fecko,fecko2}
The second source of friction is the interaction of oscillations of the
reaction co-ordinate with a modulating low frequency mode,
which in turn is strongly damped by the environment.
This is particularly important in hydrogen bonded systems.\cite{kuhn,kuhn2}
For the case, A-H$\cdots$B, the A-H 
stretch is modulated by the A$\cdots$B oscillations associated
with the hydrogen bond.
A third possible source of friction is anharmonicities\cite{Nitzan}
and Fermi resonances with other vibrational modes in the
biomolecule.

\subsubsection{Comparison with the spectral density for biological chromophores}

The dynamics of optically active molecules (chromophores) within 
proteins            have been studied extensively, both theoretically 
and experimentally.\cite{gilmore} An optical transition from the ground to
an excited electronic state is usually associated with a change in
electric dipole moment, $\Delta \mu$  of the order of a few
Debye, which then couples to the electric field (reaction field $R(t)$)
associated with the dielectric relaxation of the chromophores environment.
The relaxation can usually be assigned to three components of the 
environment: the surrounding protein, water bound to the surface of the 
protein, and the bulk water surrounding the protein.
The corresponding time scales are of the order of
nanoseconds, 10-100 psec, and 100 fsec-1 psec.
If we consider    a proton 
at the same location as the chromophore 
and described by a continuous  co-ordinate $x$,
then the time-dependent interaction energy with the reaction field
is $e x(t) R(t)$, where $e$ is the proton charge. 
For the chromophore, modelled as
a two-level system the interaction energy is
$\Delta \mu R(t) (P_e - P_g)(t)$ 
where the last factor represents the relative
occupation probability of the excited and ground
states of the chromophore.
Then the spectral
density, $J(\omega) = M {\rm Re} \ \gamma(\omega) \omega$,
relevant to the Caldeira-Leggett model
can be related to the spectral density $J_c(\omega)$
associated with a spin boson model
for the chromophore,\cite{gilmore}
\begin{equation}
J(\omega) = \left(  \frac{ e}{     \Delta \mu} \right)^2
J_c(\omega).
\label{eq:chrom}
\end{equation}
Comparison with ultra-fast laser spectroscopy experiments,
with molecular dynamics simulations, and continuum
dielectric models shows that\cite{gilmore} the high-frequency
($\omega >$ 10 cm$^{-1}$) part of the chromophore spectral
density is dominated by the bulk water.

It should be noted        that the femtosecond laser spectroscopy experiments
used to extract spectral densities do have limited time
resolution  ($\sim $ 10--100 fsec)
and so only give information about the
spectral density for $\omega < $
 500 cm$^{-1}$.
Furthermore, quantum Kramers theory requires a knowledge of $\gamma(\omega)$ and $\hat{\gamma(z)}$
at frequencies of order $\omega_0$ and $\omega_b$.

Fig. 1 of Lang et al.\cite{Lang99JCP} shows the frequency dependence of
a chromophore-water spectral density for frequencies up to 
about $\omega \sim $ 3000 cm$^{-1}$, calculated by the
method of Song and Chandler.\cite{Song98}
It has peaks of magnitude of the order of a few hundred
cm$^{-1}$, at frequencies of about 180, 600, 1800, and 3200 cm$^{-1}$. 
 There is a substantial contribution
from the librational band of water centred at 600 cm$^{-1}$.

If we combine (\ref{bound}) with (\ref{eq:chrom}) we find that
the curvature $K_e$ can be related to the reorganisation energy $E_R$
associated with the excited state of the chromophore\cite{gilmore},
\begin{equation}
K_e       =  \frac{2}{\pi} \left(  \frac{ e }{\Delta \mu} \right)^2 E_R    
\label{eq:kappae}
\end{equation}
and
\begin{equation}
\label{bound2}
\hat{\gamma}(z) \leq
\frac{2}{\pi M z}  \left(  \frac{ e }{\Delta \mu} \right)^2 E_R
\end{equation}
Given typical values of $E_R$ of $\sim 1000$ cm$^{-1}$\cite{gilmore}
and $\Delta \mu \sim $ 5 D, for chromophores in proteins
($(\hbar e/\Delta \mu)^2/M \simeq$ 30 cm$^{-1}$ for protons and  $\Delta \mu = $ 1 D)
(\ref{bound2}) then gives an estimated upper bound
\begin{equation}
\label{bound3}
\frac{\hat{\gamma}(z)}{z} \leq \sim \left(\frac{150 {\rm cm}^{-1}}{z}\right)^2.
\end{equation}
Hence, we see that for typical protein and solvent environments we expect
to be in the weak friction limit for proton transfer reactions at room temperature,
since according to the discussion around Eqn. (\ref{weak}) only values of $z \sim \omega_b$ and larger are relevant.

\subsubsection{Infra-red spectroscopy}

Infra-red (IR) spectroscopy provides a means to measure the
frequencies, damping, and decoherence of vibrational modes
in a molecule.
Recent advances in femtosecond two-dimensional IR spectroscopy has
yielded such information for several specific systems,\cite{cho}
including O-H stretches in water,\cite{woutersen}
 N-H stretches in Watson-Crick base pairs\cite{woutersen2}.
It is generally found that hydrogen bonding leads
to broad IR spectra.\cite{woutersen,kuhn,yamashita}

\subsubsection{Molecular dynamics simulations}

Stimulated by recent time-resolved infra-red spectroscopy
measurements of vibrational energy relaxation and
dephasing a number of molecular dynamics studies have been made for
specific vibrations in biomolecules.
The difficulties associated with extracting the vibrational energy relaxation rate, $1/T_1$ from
molecular dynamics simulations
has recently been summarised.\cite{fujisaki} 
If the vibration frequency is in the classical regime
($\omega_0 < k_B T \simeq 200 
\ {\rm cm}^{-1}$)
then the Landau-Teller equation allows one to extract 
$T_1$ from the classical force-force correlation function.
However, most modes of interest are not in this classical regime.
Fujisaki and  Straub\cite{fujisaki}
considered the specific case of a C-D vibration
in cytochrome-c surrounded by water.
The experimental value of $T_1$ for this mode is about one psec.
They found that the value of $T_1$ found in the simulations
 could vary by as much
as two orders of magnitude with only a ten per cent change
in the bond force constant.
Time resolved infra-red spectroscopy experiments show
that the C=O stretch of the peptide bond for a wide
range of proteins has a relaxation time of
about one psec.
A recent molecular dynamics simulation\cite{stock}
yielded values that were two orders of magnitude larger than this.

A recent combined molecular dynamics-quantum mechanics calculation\cite{marx}
calculated the IR spectrum for the water networks
in (the proton pump) bacteriorhodopsin. They found broad continuum bands,    
around 1800 and 2700 
  cm$^{-1}$, and associated them with the solvated Zundel
complex (H$_5$O$_2^+$) and Eigen complex (H$_3$O$^+$).
A time-resolved fourier-transform infrared spectroscopy
experiment found that the precise arrangement of the water
molecules within bacteriorhodopsin
was crucial to proton transfer.\cite{gerwert}
Given that many proton transfer reactions in enzymes also involve
water molecules inside the protein, such
hydrogen bonding networks may also be a significant source of friction
for proton transfer.

Moritsugu and Smith performed molecular dynamics simulations of 
myoglobin both in water and in vacuum.
They used a Langevin model to describe the dynamics of the
different vibrational models.\cite{moritsugu}
The frictional damping of different vibrational modes 
of the protein was found to be proportional
to the accessible surface area of the mode, confirming the importance
of the solvent which we have stressed here.
At 300 K, they found that for modes with $\omega$ in the range 100-400  cm$^{-1}$,
the friction could be fit to
\begin{equation}
\label{eq:smith}
{\rm Re} \ \gamma(\omega) =  
\Delta\gamma + A \omega 
\end{equation}
with $\Delta\gamma \simeq 20$
 cm$^{-1}$,
and $A \simeq 0.38$ and was temperature independent between 120 and 300 K.

\subsubsection{Dielectric continuum models}

Continuum models\cite{Hsu97,Song98,gilmore}   allow one to 
express the spectral density in terms of the frequency dependent
(complex) dielectric function $\epsilon_s(\omega)$ of the solvent.
If the proton is at the centre of a spherical cavity of
radius $a$ inside the water, and undergoes displacements
much less than the radius,         then using (\ref{eq:chrom}) and results for chromophores then
\begin{equation}
\label{eq:J(omega)}
{\rm Re} \ \gamma(\omega) =  \frac{e^2}{2\pi\epsilon_0 a^3 M \omega} {\rm Im}
\left(
 \frac{ \epsilon_s(\omega)-\epsilon_c 
}{2\epsilon_s(\omega)+\epsilon_c}
\right)
\end{equation}
 and $\epsilon_c$ is the (static) dielectric constant of the cavity,
which can be approximated as the local dielectric constant
of the protein environment surrounding the proton.

Measurements of the
frequency dependent dielectric constant of water
$\epsilon_s(\omega)$ 
in the range 1-200
  cm$^{-1}$,
have been fit to a form involving three Debye relaxation terms
and one damped resonant term\cite{Vij04JML}
\begin{equation}
\label{eq:Debye dielectric}
\epsilon_s(\omega) = \epsilon_\infty + 
\sum_{i=1}^{3}
\frac{\Delta\epsilon_i}{1+i\omega\tau_{i}}
+\frac{\Delta\epsilon_4}{1+i\omega\tau_{4}-\omega^2/\omega_4^2}
\end{equation}
where $\tau_{i}$ is the Debye relaxation time of the 
relevant component.
For water at 298 K, the co-efficients are
$\Delta \epsilon_i$ ($i=1,2,3,4$) = 71.5, 2.8, 1.6, and 0.92, respectively.
The corresponding relaxation times are
 $\tau_i$ = 8.3, 1.0, 0.1, and 0.025 psec.
The resonant frequency is
 $\omega_4=175$ cm$^{-1}$.
Roy and Bagchi\cite{roy} gave a resonant
frequency of
 $\omega_4=200$ cm$^{-1}$ and a damping constant such
that 
 $\omega_4 \tau_4=2$.
They also calculated the frequency dependent friction
for outer sphere electron transfer reactions in water
out to $\omega \tau = 5$ where $\tau=$ 0.1 psec.

Figure 5 of Ref. \onlinecite{Vij04JML} shows that for 
 $\omega \sim 100$ cm$^{-1}$ that Im$\epsilon(\omega) \sim 2$
(see also Fig. 18 in Ref. \onlinecite{franks})
which is an order of magnitude larger than the contribution
from the slowest relaxation ($i=1$).



A parametrisation of the higher frequency part of the dielectric
function has been given.\cite{nandi}
There are features
at frequencies of about 180, 600, 1800, and 3200 cm$^{-1}$. 
The first can be assigned to hindered translation
of the hydrogen-bonded network (the O$\cdots$O stretch
of the O-H$\cdots$O of the water hydrogen bonds). 
Hindered rotation (libration) is the origin of the second feature.
 Fig. 3 in Ref. \onlinecite{Hsu97} shows a plot
of the frequency dependence of the right hand site of
(\ref{eq:J(omega)}).
 Again, there is a substantial contribution
from the librational band in the range
600-800 cm$^{-1}$.

We can model a peak at a frequency $\omega_r$ in the spectral density by 
(\ref{gamma-peak}) with the corresponding friction kernel,  (\ref{gamma-peakz}).
If $z,\omega_r \gg \Gamma$ then
\begin{equation}
\hat{\gamma}(z) \sim \frac{\gamma_r \Gamma}{z}.
\end{equation}
and for the typical values of
 $\gamma_r \sim \Gamma \sim $ 100's cm$^{-1}$,
 $\omega_b \sim \nu = 2 \pi k_B T \sim$ 1000  cm$^{-1}$ this gives
$\hat{\gamma}(\nu) \ll {\nu}$ which justifies using the weak-friction
limit in reaction rate theory.

\section{Quantum Transition State Theory describes the experimental data}\label{Explain}

 Quantum Transition State Theory  predicts that
for $T > T_0$ the H/D KIE is given by
\begin{equation}\label{QTSTFit}
\frac{k_H}{k_D} = \sqrt{2}
\frac{\sinh(\hbar\omega_0/2k_B T)}{\sinh(\hbar\omega_0/2\sqrt{2}k_B T)}
\frac{\sin(\hbar\omega_b/2\sqrt{2}k_B T)}{\sin(\hbar\omega_b/2k_B T)}
 \end{equation}

\begin{figure}[ht]
\includegraphics[scale=0.7]{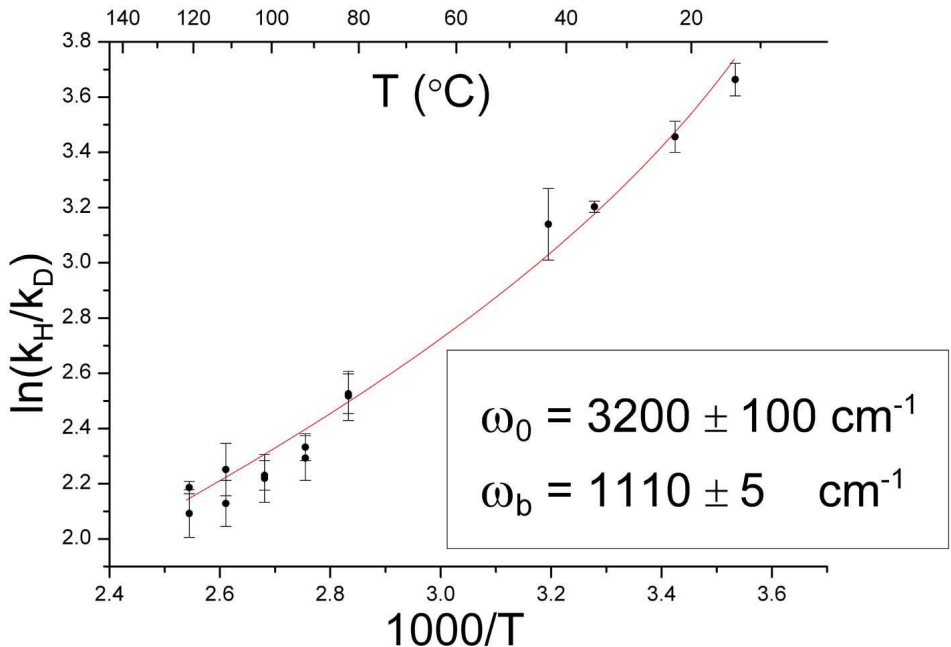}
\caption{Comparison of the  measured temperature dependence of the kinetic isotope effect\cite{Doll2003} for the enzyme methylmalonyl-CoA mutase and several synthetic molecules with similar  
H-atom abstraction reactions (NpCb, AdoCbl and 8-MeOAdoCbl)
 with Quantum Transition State Theory at temperatures above
which tunneling is possible. These systems all have KIE's and Arrhenius parameters that differ by factors of 5-10 from the semi-classical
values traditionally claimed to hold in the absence of tunneling (see Table I).
 The use of the synthetic molecules allows
coverage of a much wider temperature range (from 10 to 120$^{\circ}$C)
than possible with enzymes. The solid line is 
a fit of Eqn. (\ref{QTSTFit}) to the experimental data with two free
parameters, the barrier frequency, $\omega_b$, and the oscillation frequency
in the reactant well, $\omega_0$. 
The value obtained for $\omega_0$ is comparable with
typical C-H stretch frequencies.
The value obtained for $\omega_b$ is comparable with
 estimates  from quantum chemistry calculations (see Table \ref{tab:omegab}) of similar reactions. 
 The value of the crossover temperature $T_0 \simeq$ 250 K =  $-20 ^{\circ}$C implied by the fitted value of $\omega_b$ and Eqn. (\ref{T00}) is consistent with the domain of validity of (\ref{QTSTFit}) (i.e., $T > T_0$). 
This Figure shows that it is {\it not necessary} to invoke quantum tunneling to obtain a quantitative description of the experimental data
for this enzyme and its synthetic analogues.
}
\label{han2}
\end{figure}

\begin{figure}[ht]
\includegraphics[scale=0.7]{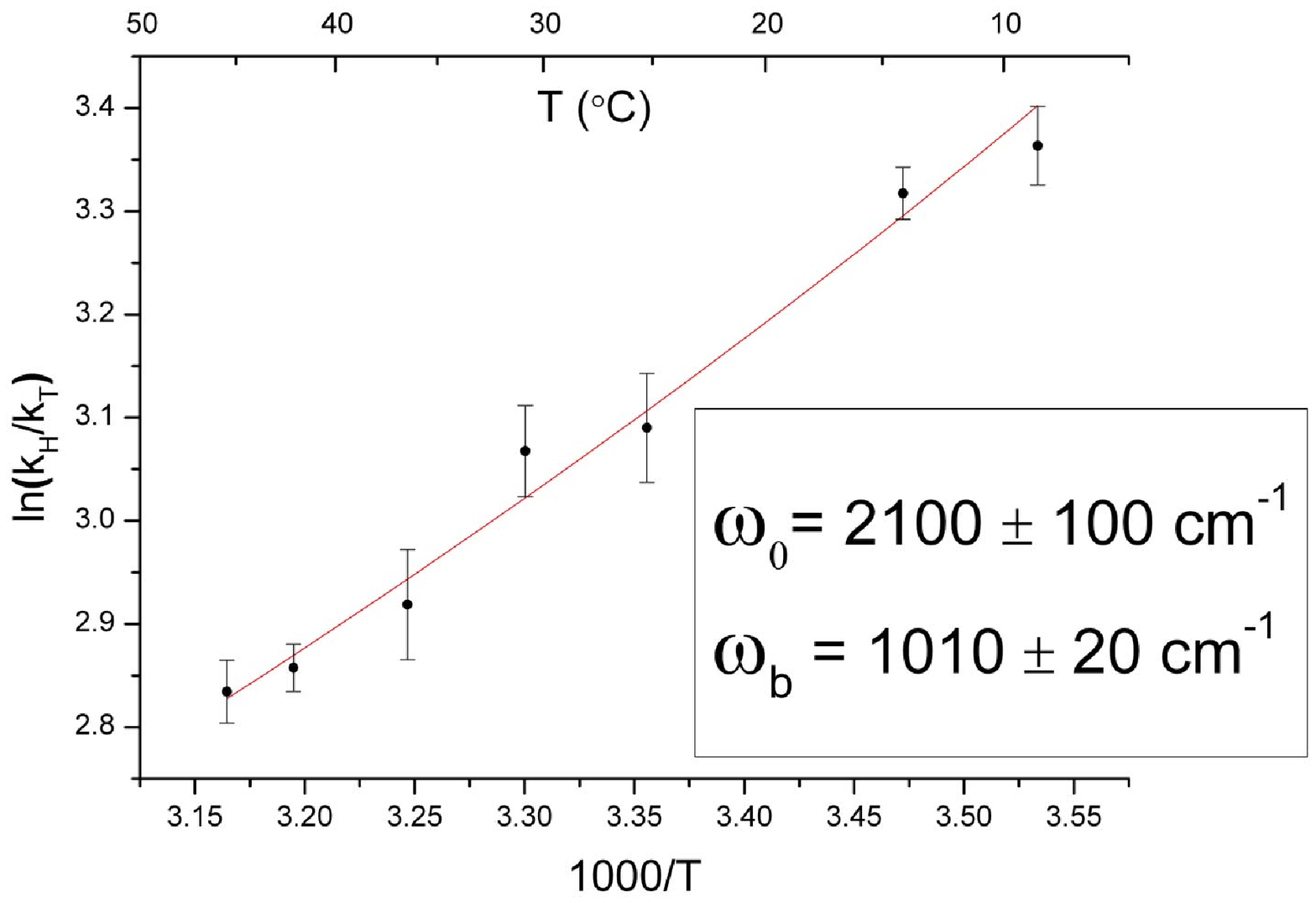}
\caption{Comparison of the temperature dependence of the kinetic isotope effect (for tritium substitution) measured for flavoenzyme monoamine oxidase
~\cite{Jonsson1994}  with Quantum Transition State Theory at temperatures above
which tunneling is possible. The solid line is 
a fit of Eqn. (\ref{eq:KIE}) to the experimental data with two free
parameters, the barrier frequency, $\omega_b$, and the oscillation frequency
in the reactant well, $\omega_0$. The 
value obtained for $\omega_b$ is comparable  with estimates  from quantum chemistry calculations (see Table \ref{tab:omegab})
for an amineoxidase enzyme which give a
barrier frequency  around 1000 cm$^{-1}$~\cite{MAOQ}. The 
value of 2100 cm$^{-1}$ for the reactant well
oscillation frequency is about two-thirds  of typical carbon hydrogen stretch
frequencies, but is comparable to values of the stretch frequency 
calculated near  transition states \cite{Cui2002,Cristobal2002}.
 The value of the crossover temperature $T_0 \simeq$ 240 K implied by the fitted value of $\omega_b$ and Eqn. (\ref{T00}) is consistent with the domain of validity of (\ref{QTSTFit}). 
This Figure shows that it is {\it not necessary} to invoke quantum tunneling to obtain a quantitative description of the experimental data
for this enzyme.
}
\label{han1}
\end{figure}

Note that this expression depends only on two parameters, $\omega_0$ and $\omega_b$.
  Figures \ref{han2} and \ref{han1} show the fits of Eq. \eqref{QTSTFit} to experimental data for different enzymes
 that show kinetic anomalies typical of systems that have
been argued to exhibit          tunneling. 
The first thing to notice is that the quantum transition state theory result can reproduce these experimental results. Secondly, the values of $\omega_0$ and $\omega_b$ obtained from
the fits are comparable to typical C-H stretch frequencies 
 and to barrier frequencies that have been obtained from quantum chemistry calculations (Table \ref{tab:omegab}). The 
values of $\omega_b$ obtained for both fits imply a crossover 
temperature $T_0$ below room temperature,
indicating that our description is  self-consistent
 in that we are in the temperature regime above the crossover.

%

\section{Conclusions}

The path integral approach we have used has several benefits for elucidating
questions concerning the role of quantum tunneling in hydrogen transfer reactions
in enzymes. It allows a full quantum mechanical treatment of the role of both temperature and the environment of the active site.
Quantum tunneling is described by the instantons (bounce solution),
 which are periodic
solutions to the semi-classical equations of motion in imaginary time. An
important result is that these solutions only exist below some temperature,
$T_0$ which is determined by the curvature of the top of the barrier.
Above this temperature the only role of quantum effects concerns quantum
fluctuations about the transition state.
We have shown that for two enzymes a quantitative description of
the temperature dependence of kinetic isotope effects 
is possible in terms of such a quantum transition state theory.
This suggests that quantum tunneling does not play as
significant role in hydrogen transfer as is often claimed.

\section{Acknowledgements}

We thank P. Curmi, P. Davies, H. Grabert, 
N. Hush, M. Karplus, J. Klinman, H.A. McKenzie, 
P. Meredith, S. Olsen, B.J. Powell, J. Reimers, H.F. Schaefer,
M.F. Smith, and W. Yang  for helpful discussions. This work was supported by the Australian Research Council.

\begin{appendix}
\section{The Swain-Schaad exponent is an unreliable
 criteria for quantum tunneling}

 A parameter that has been used to quantify the effect of isotope substitution is the Swain-Schaad exponent, which is defined as 
\begin{equation}\label{swaa}
\alpha=\frac{\ln(k_H/k_T)}{\ln(k_D/k_T)}.
\end{equation}

This parameter is used because in the standard approach to reaction rate theory the Swain-Schaad exponent is independent of temperature and system specifics~\cite{swaa}. It  can be written as
\begin{equation}
\alpha=
 \frac{\ln(A_H/A_T)-\left[E_P - E_T\right]/k_B T }{\ln(A_D/A_T)-\left[E_D - E_T\right]/k_B T  }.
\end{equation}
In the limit of low temperatures 
and where  $A_H/A_T \hspace{1.5mm} $ and $ \hspace{1.5mm} A_D/A_T \to 1$
 this simplifies to 
\begin{equation}
\frac{\ln(k_H/k_T)}{\ln(k_D/k_T)}= \frac{E_P - E_T}{E_D - E_T}
\simeq 3.3
\end{equation}

From the standard semi-classical rate theory the ratio of the differences in activation energies of the different isotopes is a constant that does not depend on the temperature or system specifics.  In this case the value that $\alpha$
 takes will depend on the prefactor values and also on temperature. This subtlety has sometimes been overlooked in the experimental literature. There are a number of cases where a discrepancy between the theoretical value of $(E_P - E_T)/(E_D - E_T)$ and the experimentally determined value of $\ln(k_H/k_T)/\ln(k_D/k_T)$ have formed the basis for the hypothesis that tunneling is occurring, even though all the terms in the above expression
are comparable\cite{Bahnson1993,Cha1989Science,Grant1989}. It is for this reason that in this paper we do not use the Swain-Schaad exponent  
as a criteria for the presence of tunneling.

\section{Effect of the environment on the crossover temperature}

\begin{figure}[htp]
\centering
\includegraphics[scale=0.8]{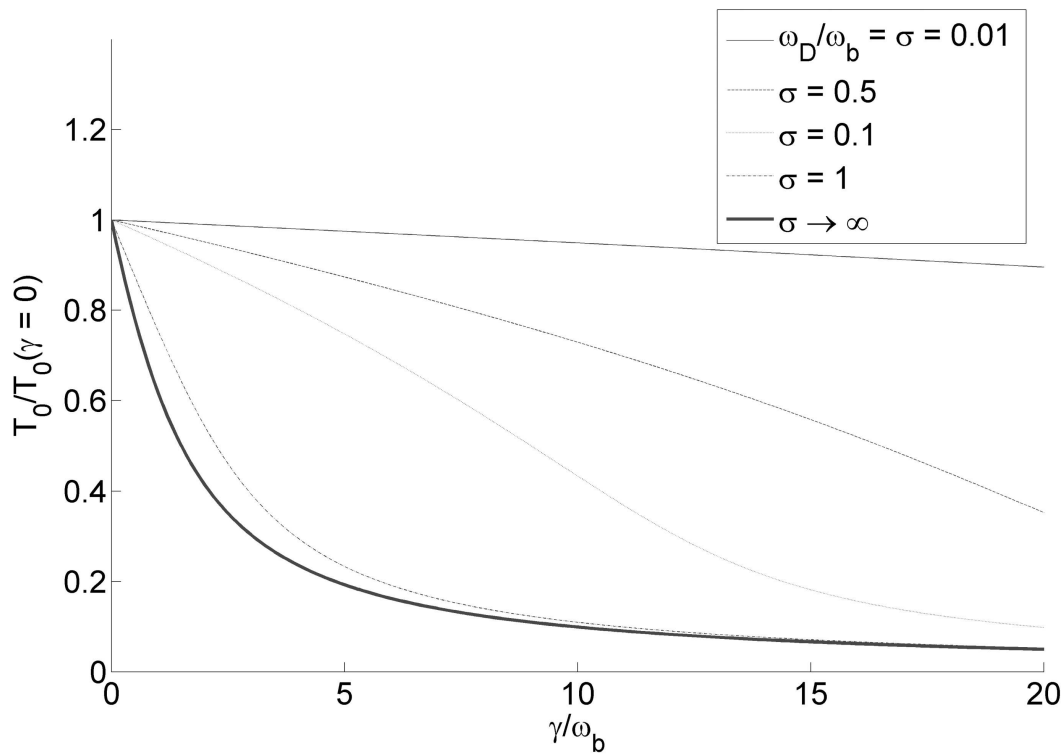}
\caption{The crossover temperature for a parabolic
barrier as a function of the friction strength for a Lorentzian spectral density. The friction strength is the dimensionless parameter $\gamma/\omega_b$. The different plots show how the relative size of the response frequency of the bath, $\omega_D$, and the barrier frequency, $\omega_b$ change the influence that friction has on the crossover temperature. These show that when  $\omega_D \gg \omega_b$ the crossover temperature becomes most sensitive to the friction strength. Then                                the bath is able to respond on a timescale which is much faster than all of the other relevant timescales.  When the response frequency of the bath is much less than the barrier frequency the crossover temperature is only weakly suppressed as the friction strength increases.}
\label{T0fri}
\end{figure}

In the case of a Lorentzian spectral density equation for $\mu$ takes the form
\begin{equation}\label{od}
\mu^2-\omega_b^2+\frac{\mu\omega_D\gamma}{\omega_D+\mu}=0.
\end{equation}
 $T_0$ varies as a function of the friction strength for different bath response frequencies, $\omega_D$. Fig. \ref{T0fri} shows how the positive root of Eq. \eqref{od} changes as a function of the scaled friction and bath response frequency.
For all values of $\omega_D$ an increase in the strength of the damping, $\gamma$, reduces the effective barrier frequency and hence the crossover temperature.

\end{appendix}

\bibliography{Hydrotun}

\end{document}